\documentclass[floatfix,epsf,prb,twocolumn,amsmath,amssymb,showpacs,superscriptaddress]{revtex4}
\usepackage[T1]{fontenc}
\usepackage[english]{babel}
\usepackage{graphics}
\usepackage{amsmath}
\usepackage{amsfonts}
\usepackage{dsfont}
\usepackage{braket}
\usepackage{graphicx}
\usepackage{subfigure}
\usepackage{color}

\renewcommand{\Re}{\mathrm{Re}\,}
\renewcommand{\Im}{\mathrm{Im}\,}
\renewcommand{\i}{\mathrm{i}}
\newcommand{\e}{\mathrm{e}}
\newcommand{\eps}{\varepsilon}
\newcommand{\p}{\partial}

\newcommand{\ph}{\varphi}

\begin{document}
\title{Scattering of a Dirac electron on a mass barrier}

\author{A.~Matulis}\email{amatulis@takas.lt}
\affiliation{Departement Fysica, Universiteit Antwerpen \\
Groenenborgerlaan 171, B-2020 Antwerpen, Belgium}
\affiliation{Semiconductor Physics Institute, Center of Physical
Sciences and Technology,\\
 Go\v{s}tauto 11, LT-01108 Vilnius, Lithuania}
\author{M.~Ramezani Masir}\email{mrmphys@gmail.com}
\affiliation{Departement Fysica, Universiteit Antwerpen \\
Groenenborgerlaan 171, B-2020 Antwerpen, Belgium}
\author{F.~M.~Peeters}\email{francois.peeters@ua.ac.be}
\affiliation{Departement Fysica, Universiteit Antwerpen \\
Groenenborgerlaan 171, B-2020 Antwerpen, Belgium}

\begin{abstract}
The interaction of a wave packet (and in particular the wave front) with a
mass barrier is investigated in one dimension.
We discuss the main features of the wave packet that are inherent to
two-dimensional wave packets, such as compression during
reflection, penetration in the case when the energy is lower than
the height of the barrier, waving tails, precursors and the
retardation of the reflected and penetrated wave packets.
These features depend on the wave packet envelope function
which we demonstrate by considering the case of a rectangular wave packet with
sharp front and trailing edges and a smooth Gaussian
wave packet. The method of Fourier integral for obtaining the
non-stationary solutions is used.
\end{abstract}

\pacs{03.65.Pm, 73.22.-f, 73.63.Fg}

\maketitle

\section{Introduction}

Progress in nanotechnology has triggered a broad interest in
low dimensional physics. The separation of graphene by mechanical exfoliation in 2004
\cite{R1,R2} opened a broad range of activity for theoretical
and experimental researchers. The interest rose due to the
peculiar properties of graphene such as ultra-relativistic
behavior of charge carriers with a Fermi velocity $300$ times
smaller than the velocity of light, the linear spectrum close to
$K$ and $K'$ points in the momentum plane that can be described
by the massless Dirac-Weyl equation\cite{R1,R3}, an unconventional
quantum Hall effect \cite{R4}, and the perfect transmission
through arbitrarily high and wide barriers, the so called Klein
tunneling \cite{R5}.

The energy levels or the spectrum of the electron system typically
investigated as being the main characteristics of quantum
nano-structures (say, such as quantum dots) are found by considering
the stationary Schr\"{o}dinger equation \cite{R7}. Recently, a
scientific interest has shifted to the investigation of quantum
dynamics including quasi-bound states \cite{R8}, electron beams,
wave packets, and their control by means of barriers and other
nonhomogeneous structures \cite{R11,R24,R13}. Unfortunately due to above
mentioned Klein effect and the gapless spectrum the control of
electrons in graphene by means of electric fields is inefficient.
Therefore, the search for other possibilities to control electrons
in graphene become of interest. Recently, the creation of a gap in
the electron spectrum \cite{MF1} as well as the control of the
valley isospin \cite{mass3,mmp11} was predicted by introducing a
mass term into the Dirac-Weyl Hamiltonian. This possibility was
experimentally demonstrated by the proper arrangement of dopants in
the graphene sheet \cite{mass1} or by inducing electron-electron
interactions \cite{Max1}. Therefore, the investigation of the
non-stationary solutions of the Dirac equation describing the
interaction of electrons with mass barriers becomes timely. Such
sophisticated systems and problems are a challenge for numerical
simulation. But in order to obtain a physical understanding of the
behavior of such systems it is very helpful to invent and analyze
simple models that are able to demonstrate the main physical
features of interest.

The purpose of the present work is to demonstrate the mean properties
of wave packets, such as their reflection and penetration into mass
steps and barriers, the appearance of precursors, the formation of
evanescent waves and the zitterbewegung, using very simple analytically
solvable one-dimensional (1D) models. Comparing the results obtained for
a rectangular wave packet with the ones having a rather soft Gaussian
shaped wave packet we demonstrate that their propagation depends on the
shape of the wave packet. In our analysis we pay special attention to the
appearance of so called precursors that were predicted by Sommerfeld nearly
a hundred years ago for electromagnetic waves travelling in a dispersive
media \cite{R14}. We also focus on the waving after-effects that can be
related to the zitterbewegung that was predicted by Schr\"{o}dinger for
relativistic electrons \cite{R15}, and which we predict should be observable
in graphene \cite{R21}.
The results were obtained applying the Fourier integral technique for solving
the non-stationary Dirac equation, and analyzing the solution by means of
integration in the complex wave vector plane.

The paper is organized as follows. In Sec.~II we introduce the problem and
the method of its solution considering the reflection of the Dirac wave packet
from a hard wall. In Sec.~III results for scattering of a plane wave type
wave function by the mass barrier is discussed which enables us to present
the non-stationary solutions for the reflected front in the form of a complex
integral in Sec.~IV. The penetration of the front of the wave packet into
the barrier is considered in Sec.~V, and in Sec.~VI the scattering
of the rectangular wave packet is studied. Sec.~VII is devoted to the
description of the scattering of a Gaussian wave packet. Our conclusions are
presented in Sec.~VIII.

\section{Reflection of a wave packet from a hard wall}
\label{sec_hw}

We consider the motion of a 1D wave packet that is described by
the following Dirac-Weyl equation:
\begin{equation}\label{dir}
  i\frac{\p}{\p t}\Psi(x,t) = -i\sigma_x\frac{\p}{\p x}\Psi(x,t)
  + \Theta(x)\sigma_z\Psi(x,t),
\end{equation}
where $\sigma_x$ and $\sigma_z$ are the Pauli matrices, and the
symbol $\Theta(x)$ stands for the Heaviside step function whose
value is zero for negative argument and unity for positive one.
The last term will be referred to as the mass term, that leads to a
gap in the electron spectrum close to the $K$ and $K'$ points.

In order to simplify the notations the above equation, the results are presented in dimensionless notations
based on the height of the mass barrier $V$. So, the energy
is measured in $V$ units, the time --- in $\hbar/V$ units, and
the unit of length is $\hbar v_F/V$, where the symbol $v_F$ stands
for the Fermi velocity.
For the sake of illustration we took $V=53$ meV, what is achieved
by depositing the single graphene layer on a BN substrate (see Ref.~\onlinecite{Giov07}).
Then the time unit is a hundredth of ps, and the unit of length is about
10 nm.

Presenting the wave function as a two component spinor
\begin{equation}\label{wfcomp}
  \Psi(x,t) = \left(\begin{array}{c} u(x,t) \\ v(x,t) \end{array}\right)
\end{equation}
we have to solve the following set of two differential
equations for the wave function components:
\begin{subequations}\label{system}
\begin{eqnarray}
\label{system1}
  u_t &=& -v_x - i \Theta(x)u , \\
\label{system2}
  v_t &=& -u_x + i \Theta(x)v.
\end{eqnarray}
\end{subequations}
Apart of the wave function components themselves the wave packet can be
characterized by the local density
\begin{equation}\label{dens}
  \rho(x,t) = |u(x,t)|^2 + |v(x,t)|^2,
\end{equation}
the local current
\begin{equation}\label{curr}
  j(x,t) = 2\Re [u^*(x,t)v(x,t)],
\end{equation}
and some averaged values: the number of particles in the wave packet
(the norm of the wave function)
\begin{equation}\label{number}
  N(t) = \int_{-\infty}^{\infty} dx \rho(x,t),
\end{equation}
the mean position
\begin{equation}\label{coord}
  X(t) = N^{-1}(t)\int_{-\infty}^{\infty} dx x\rho(x,t),
\end{equation}
the mean velocity
\begin{equation}\label{velo}
  V(t) = N^{-1}(t)\int_{-\infty}^{\infty} dx j(x,t),
\end{equation}
and the width of the wave packet
\begin{equation}\label{width}
  w(t) = N^{-1}(t)\int_{-\infty}^{\infty} dx \left[x-X(t)\right]^2\rho(x,t).
\end{equation}

In order to introduce the necessary definitions and illustrate the
method of solving the time dependent Dirac equation we start with
the most simple problem: the reflection of the wave packet from a
hard wall. This means that
Eqs.~(\ref{system}) without the last term (i.~e.~the massless
Dirac equation) will be solved in the negative part of the $x$-axis
($-\infty < x \leqslant 0$) and the hard wall will be taken into
account by the boundary condition
\begin{equation}\label{bc}
  u(0,t) + iv(0,t) = 0
\end{equation}
which was derived and discussed in Ref.~\onlinecite{mmp11}.

This solution can be easily composed of two (incident and
reflected) freely propagating wave packets. The incident wave packet moving
to the right side can be presented as
\begin{equation}\label{wffree}
  \Psi^{(\mathrm{in})}(x,t) = \frac{1}{\sqrt{2}}
  \left(
  \begin{array}{c}
  1 \\
  1 \\
  \end{array}
  \right)
  \e^{iq(x - t)}\Phi(t - x),
\end{equation}
that satisfies the above massless equation with any envelope
function $\Phi(x)$. In order to satisfy the boundary condition
(\ref{bc}) at any time $t$ we have to add the reflected wave packet
\begin{equation}\label{wfrefl}
  \Psi^{(\mathrm{rfl})}(x,t) =
  \frac{-i}{\sqrt{2}}\left(
  \begin{array}{c}
  1 \\
  -1 \\
  \end{array}
  \right)
  \e^{-iq(x+t)}\Phi(t + x),
\end{equation}
that propagates to the left.

The most interesting feature of the wave function constructed in
such a way
\begin{equation}\label{wftot}
  \Psi(x,t) = \Psi^{(\mathrm{in})}(x,t) + \Psi^{(\mathrm{rfl})}(x,t),
\end{equation}
and describing the Dirac electron reflection from the hard wall, is
the absence of interference of the incident and reflected wave packets.
Indeed, denoting
\begin{equation}\label{defcomp}
  f_{\pm} = \e^{iq(\pm x - t)}\Phi(t\mp x),
\end{equation}
we present the density as
\begin{equation}\label{dens0}
\begin{split}
\rho(x,t) &= |f_+ - if_-|^2 + |f_+ + if_-|^2\\
  &= |f_+|^2 + |f_-|^2 = |\Phi(t-x)|^2 + |\Phi(t+x)|^2,
  \end{split}
\end{equation}
and the current as
\begin{eqnarray}\label{curr0}
\begin{split}
j(x,t) &= \Re\left\{\left(f_+^* + if_-^*\right)\left(f_+ + i f_-\right)\right\}\\
  &= \Re\left\{|f^+|^2 - |f^-|^2 + 2\i\Re\left(f_+^*f_-\right)\right\}\\
  &= |f^+|^2 - |f^-|^2 = |\Phi(t-x)|^2 - |\Phi(t+x)|^2.
\end{split}
\end{eqnarray}
Thus, the density and current are expressed just through the
corresponding individual values of the incident and reflected
wave packets without any interference terms.

In order to trace how the reflection depends on the form of the
envelope function we consider in detail two extreme shapes of wave
packets: a wave packet
with a rectangular envelope as an example of a wave packet with
abrupt edges, and a rather soft Gaussian one.

\begin{figure}[ht]
\begin{center}
\includegraphics[width=8cm]{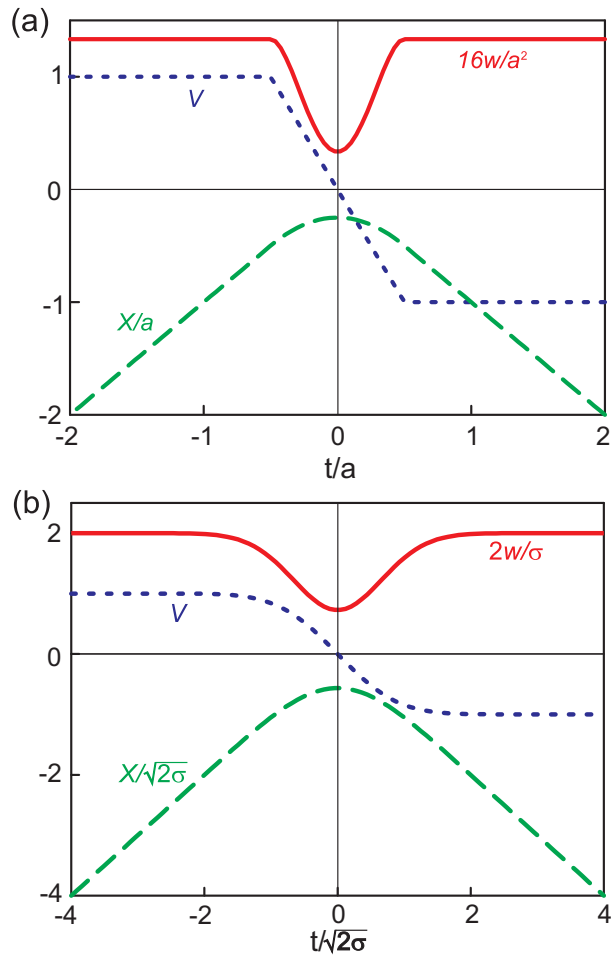}
\caption{(Color online) Evolution of mean values for the Dirac wave packet
  in time: the mean coordinate of the wave packet -- green dashed curve,
  the mean velocity -- blue dotted curve and the width -- red solid curve.
  (a) -- the rectangular wave packet with envelope (\ref{envrect}) and
  (b) -- the Gaussian one with envelope (\ref{gpbc}).} \label{fig1}
\end{center}
\end{figure}

We choose the following envelope for the rectangular wave packet:
\begin{equation}\label{envrect}
  \Phi(x) = \frac{1}{a}\Theta(a^2/4 - x^2).
\end{equation}
Inserting this into (\ref{coord}--\ref{width}) we obtain the
following mean values
\begin{subequations}\label{meaninf}
\begin{eqnarray}
  X(t) &=& \displaystyle{-|t|\Theta\left(t^2-\frac{a^2}{4}\right)
  - \frac{4t^2+a^2}{4a}\Theta\left(\frac{a^2}{4}-t^2\right)}, \phantom{mmm} \\
  V(t) &=& \displaystyle{-\frac{t}{|t|}\Theta\left(t^2-\frac{a^2}{4}\right)
  -\frac{2t}{a}\Theta\left(\frac{a^2}{4}-t^2\right)}, \\
  w(t) &=& \frac{a^2}{12} -\left(\frac{4t^2 - a^2}{4a}\right)^2
  \Theta\left(\frac{a^2}{4}-t^2\right).
\end{eqnarray}
\end{subequations}
characterizing the reflection of the rectangular wave packet from the
hard wall. They are shown in Fig.~\ref{fig1}(a).
We would like to draw your attention to the following features of
this simple example of reflection: (1) the wave packet is reflected
before it reaches the wall, (2) it is compressed during
the reflection, and (3) the wave packet dimensions are restored after
the reflection. This behavior can be compared with the one of a rubber ball hitting
a hard wall: due to its finite extension the ball changes the direction of motion
before its center reaches the wall and it is squeezed during the collision.

In the case of a Gaussian wave packet the following envelope function was chosen:
\begin{equation}\label{gpbc}
  \Phi(x) = \frac{\e^{iqx}}{(2\pi\sigma)^{1/4}}\e^{-x^2/4\sigma},
\end{equation}
where the parameter $\sigma$ characterizes the width of the wave packet
and plays the same role as the parameter $a$ in the previous case
of a rectangular wave packet. Now inserting this expression into Eqs.~%
(\ref{wffree}--\ref{wftot}) and then into Eqs.~(\ref{coord}--\ref{width})
we obtain the following averaged values:
\begin{subequations}\label{xvwgauss}
\begin{eqnarray}
  X(t) &=& -\sqrt{\frac{2\sigma}{\pi}}\e^{-t^2/2\sigma}
  - t\cdot\mathrm{erf}(t/\sqrt{2\sigma}), \\
  V(t) &=& -\mathrm{erf}(t/\sqrt{2\sigma}), \\
  w(t) &=& \sigma + t^2 - \left[\sqrt{\frac{2\sigma}{\pi}}\e^{-t^2/2\sigma}
  - t\cdot\mathrm{erf}(t/\sqrt{2\sigma})\right]^2\phantom{mm}
\end{eqnarray}
\end{subequations}
where the symbol "$\mathrm{erf}$" stands for the error function.
These dependencies  are shown in figure Fig.~\ref{fig1}(b).
Comparing both Figs.~(a) and (b) we see that qualitatively they are the same,
and consequently, the reflection of the Dirac wave packet from the infinite wall
is not sensitive to the form of the envelope function. This non-sensitivity
is quite expected because the dispersionless propagation of the wave packet
described by Eq.~(\ref{dir}) without the mass term is not spoiled by the
dispersionless boundary condition (\ref{bc}).
\section{Mass barrier of a finite height}
\label{sec_exp}

Now we consider our main problem: the interaction of a wave
packet with a mass barrier of finite height. This problem is
much more complicated than the previous one because now the wave packet
spends some time inside the barrier region which is a dispersive
medium, and as a consequence the wave packet will not conserve its shape.

As the Dirac equation is linear with coordinate independent
coefficients a natural way to solve the problem is by using Fourier
transformation. Thus, we choose the initial condition (say the position
of the wave packet at time $t=0$), expand it into Fourier series (here integral
over momentum $k$), and then change the exponent in the integrand
by the function that obeys the time dependent Dirac equation in
both regions (the barrier and free motion region) and is
consistent with the boundary conditions at the point $x=0$.
For this purpose we recap briefly the main results for
scattering of a plane-wave (exponent) type wave function.

We assume that in the region of free motion ($-\infty<x<0$) there
are incident and reflected waves with energy $\eps=k$, and here
the total wave function reads
\begin{equation}\label{increfl}
\begin{split}
\Psi^{(\mathrm{free})}(x,t) &= \Psi^{(\mathrm{in})}(x,t) + \Psi^{(\mathrm{rfl)}}(x,t)\\
  &=\frac{1}{\sqrt{2}}\left\{\left(
                              \begin{array}{c}
                                1 \\
                                1 \\
                              \end{array}
                            \right)
  \e^{ikx} + R\left(
                  \begin{array}{c}
                    1 \\
                    -1 \\
                  \end{array}
                \right)
  \e^{-ikx}\right\}\e^{-ikt}.
  \end{split}
\end{equation}
In the barrier region ($0<x<\infty$) there is only an outgoing wave
moving to the right with the same energy $k$ and momentum $\kappa
= \sqrt{k^2 - 1}$. We present its wave function as
\begin{equation}\label{outgo}
  \Psi^{(\mathrm{tr})}(x,t) =
  \frac{T}{\sqrt{2k}}\left(
  \begin{array}{c}
  \sqrt{k+1} \\
  \sqrt{k-1} \\
  \end{array}
  \right)
  \e^{i\kappa x}\e^{-ikt}.
\end{equation}

Now equating both wave function components at the point $x=0$ we
obtain the following wave reflection and penetration amplitudes:
\begin{subequations}\label{rtexp}
\begin{eqnarray}
\label{rtexp1}
  R(k) &=& k - \kappa, \\
\label{rtexp2}
  T(k) &=& \sqrt{k}\left(\sqrt{k+1} - \sqrt{k-1}\right).
\end{eqnarray}
\end{subequations}

Eqs.~(\ref{increfl}) and (\ref{outgo}) together with
definitions (\ref{rtexp}) enable us to present the time dependent
wave function of any wave packet in the form of the integral along some
contour in the complex $k$-plane.

\section{Reflection of a steep front}
\label{sec_rsf}

First we consider the motion of the wave packet with rectangular
envelope function. Fortunately, due to the linearity of the Dirac
equation this problem can be decomposed into the motion of two
fronts corresponding to the leading and trailing edges of this
wave packet. That is why we start with the reflection
of a steep front from the mass barrier, and choose the initial
incident wave packet as
\begin{equation}\label{initinc}
  \Psi^{(\mathrm{inc})}(x,0) =
  \frac{1}{\sqrt{2}}\left(
  \begin{array}{c}
  1 \\
  1 \\
  \end{array}
  \right)
  \e^{iqx}\Theta(-x).
\end{equation}
It coincides with the freely propagating front where the leading edge
has reached the barrier. This function can be presented by the
following Fourier integral:
\begin{equation}\label{incfourier}
  \Psi^{(\mathrm{inc})}(x,0) = \frac{1}{\sqrt{2}}\left(
  \begin{array}{c}
  1 \\
  1 \\
  \end{array}
  \right)
  \frac{1}{2\pi}\int_{-\infty}^{\infty}dk\e^{ikx}f(k),
\end{equation}
where
\begin{equation}\label{ff}
  f(k) = \int_{-\infty}^0 dx\e^{[i(q-k) + \alpha]x}
  = \frac{1}{i(q-k) + \alpha}.
\end{equation}
Here the symbol $\alpha$ stands for the regularization parameter
--- a small positive value that will be set to zero at the end of
the calculation.

Now according to our strategy we replace the exponent $\exp(ikx)$
in the integrand of the Fourier integral (\ref{incfourier}) by
the solution of the time dependent Dirac equation
(\ref{increfl},\ref{outgo}) describing the reflection and
penetration into the barrier of this exponent type wave function.
The wave function of the wave packet obtained in this way consists of
three parts. Two of them are defined in the free motion region
($-\infty<x<0$) and describe the incident front
\begin{equation}\label{incpack}
  \Psi^{(\mathrm{in})}(x,t) = -\frac{1}{\sqrt{2}}\left(
  \begin{array}{c}
  1 \\
  1 \\
  \end{array}
  \right)
  \frac{1}{2\pi i}\int_C\frac{dk\e^{ik(x - t)}}{k - q + i\alpha},
\end{equation}
and the reflected one
\begin{equation}\label{reflpack}
\begin{split}
&  \Psi^{(\mathrm{rfl})}(x,t) \\
&  = -\frac{1}{\sqrt{2}}\left(
  \begin{array}{c}
  1 \\
  -1 \\
  \end{array}
  \right)
  \frac{1}{2\pi i}\int_C
  \frac{\e^{-ik(x + t)}(k - \kappa)dk}{k - q + i\alpha}.
\end{split}
\end{equation}
The third part of the wave function is defined in the
region $0 < x < \infty$ and describes the front that penetrates the barrier:
\begin{equation}\label{penpack}
\begin{split}
&  \Psi^{(\mathrm{tr})}(x,t) \\
&  = -\frac{1}{2^{3/2}\pi i}
  \int_C \frac{\e^{ i(\kappa x - kt)}dk}{k - q +  i\alpha}\left(
  \begin{array}{c}
  k + 1 -\kappa \\
  \kappa - k + 1 \\
  \end{array}
  \right).
\end{split}
\end{equation}

Integral (\ref{incpack}) is trivial because its integrand has
just a single singular point in the lower part of the complex
$k$-plane, namely, the pole at the point $k=q- i\alpha$. Thus,
choosing the integration contour $C$ by passing that point from above and
enclosing it by the upper or lower semi-circle (depending on the
sign of the parameter $x-t$ in the exponent) we calculate the residue
at this pole and obtain the following wave function of the
incident front:
\begin{equation}\label{krfr}
  \Psi^{(\mathrm{in})}(x,t) = \frac{1}{\sqrt{2}}\left(
  \begin{array}{c}
  1 \\
  1 \\
  \end{array}
  \right)
  \Theta(t-x)\e^{iq(x - t)}
\end{equation}
that describes the motion of this front with a constant velocity equal to unity
due to the absence of dispersion in the free motion region.

\begin{figure}[ht]
\begin{center}
\includegraphics[width=6cm]{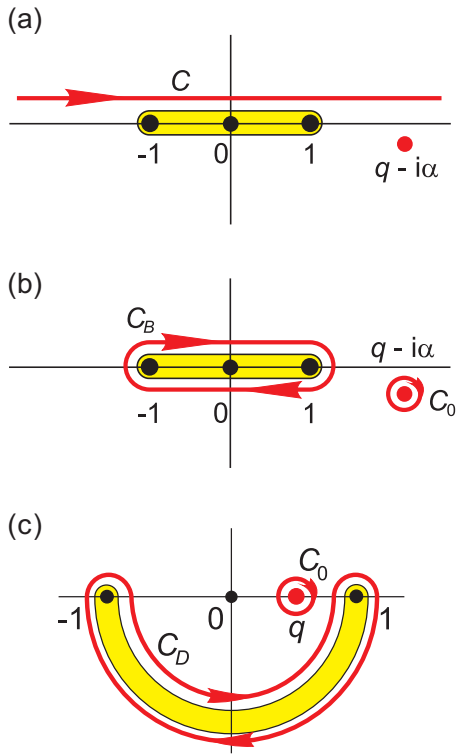}
\caption{(Color online) The complex $k$-plane, pole, cut and contour for the
  integration in equations (\ref{reflpack}) and (\ref{penpack}):
  (a) -- initial contour, (b) -- transformation of the contour
  in the case $q>1$ and (c) -- in the case $q<1$.} \label{fig2}
\end{center}
\end{figure}

The two other integrals (\ref{reflpack}) and (\ref{penpack}) are
more complicated because of the radicals $\kappa=\sqrt{k^2-1}$ in
their integrands leading to branching points at $k_{\pm}=\pm 1$ in
the complex $k$-plane. Consequently, we have to make a cut in this
plane and choose the position of the integration contour
accordingly. How this is done by taking the causality principle
into account is shown in Fig.~\ref{fig2}(a).
The choice of the Riemann sheet is defined by the requirement
$\Im\kappa>0$ in the upper half-plane, and this is determined by
the fact that at large momentum $k$ (positive or negative) the
value of $\kappa$ should approximately coincide with the value of
$k$. The choice of the integration contour $C$ laying above both
singularities (pole and cut) is in agreement with the requirement
that the integral (\ref{reflpack}) should be zero in the case of
$x<-t$ as the reflected front can not move faster than with unit
velocity.

In the case of $x>-t$ the argument of the exponent has opposite sign
and that is why the contour can be enclosed by the lower
semi-circle and transformed into two contours encircling the
singularities as it is shown in Fig.~\ref{fig2}(b,c), corresponding to the
contributions of pole and cut.

The pole contribution is calculated by means of the residue technique and it gives
\begin{equation}\label{rfpol}
  \Phi_{\mathrm{pole}}^{(\mathrm{rfl})}(x,t) = \Theta(x+t)R(q)\e^{-iq(x + t)}
  \frac{1}{\sqrt{2}}\left(
  \begin{array}{c}
  1 \\
  -1 \\
  \end{array}
  \right).
\end{equation}
It coincides with the result presented in section \ref{sec_exp} for
scattering of a plane wave. The single difference is that now
the exponent has a steep leading edge.

The calculation of the cut contribution is more complicated. It can not be
calculated analytically, and a numerical evaluation of the integral is necessary.
This integral depends essentially on the energy
$\eps= q$ of the incident front, namely, whether it is larger or
smaller than the height of the barrier.

\subsection{Above the barrier reflection ($q>1$)}

For $q>1$ the contours $C_0$ and $C_B$ are
separated horizontally (see in Fig.~\ref{fig2}(b))
and the pole of the integrand doesn't
complicate the calculation of the cut contribution along the
contour $C_B$. That is why we transformed this contribution into two real
integrals that depend on the single argument $\xi=x+t$:
\begin{equation}\label{rfpj1}
\begin{split}
&  \Phi_{\mathrm{cut}}^{(\mathrm{rfl})}(x,t) \equiv \Phi_{\mathrm{cut}}^{(\mathrm{rfl})}(\xi)\\
  &= -\frac{2}{\pi}\int_0^1 dk\frac{\sqrt{1-k^2}}{q^2-k^2}
  \left\{-q\cos(k\xi)+i\sin(k\xi)\right\}
\end{split}
\end{equation}
and integrated this numerically.

\begin{figure}[ht]
\begin{center}
\includegraphics[width=8cm]{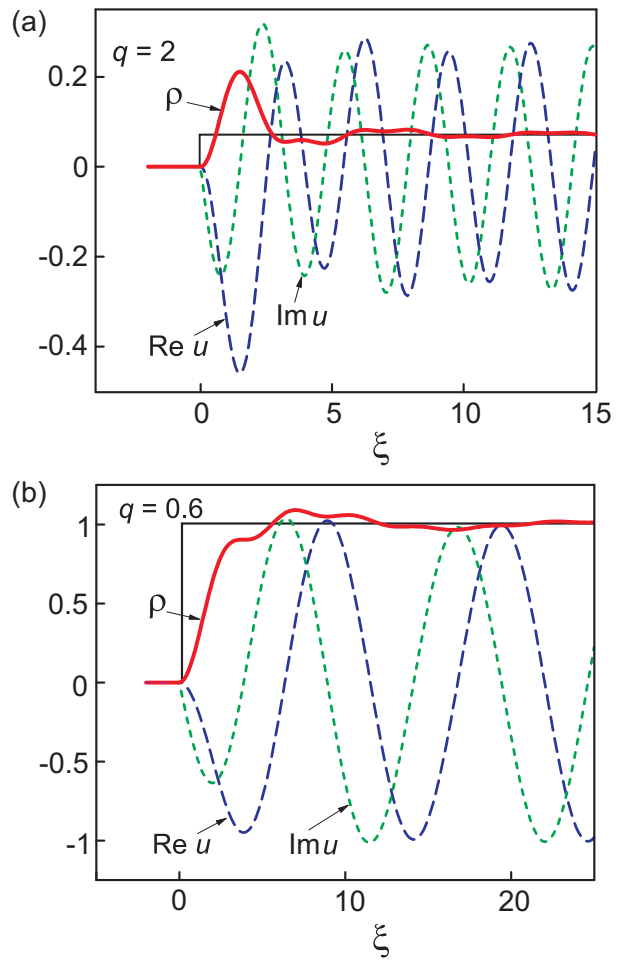}
\caption{(Color online) The wave function of the reflected front as a function of the variable $\xi=x+t$:
  blue dashed curve -- the real part of the component $u=-v$, green dotted curve
  -- the imaginary part of it, red thick solid curve -- the local density,
  and black thin solid curve -- the same local density with only the pole contribution
  included; (a) -- above the barrier reflection with energy $q=2$ and (b) --
  below the barrier reflection with $q=0.6$.} \label{fig3}
\end{center}
\end{figure}

A typical result is shown in Fig.~\ref{fig3}(a).
There are two points worth to be mentioned. First, at the beginning of
the front there is some overshoot of the wave function modulus
squared (the local density) as compared with the one when only
the pole contribution is taken into account (as was already
mentioned the pole contribution coincides with the result
of plane wave scattering).
Second, the above mentioned local density tends to the pole
contribution with increasing $\xi=x+t$ (when we are going
away from the leading edge of the front) but notice that some oscillations
are still present. Following \cite{R13} the overshoot can be explained as follows.
According to Fourier transform (\ref{incpack}) the incident front can be
considered as a superposition of numerous plane waves with energies larger
and smaller than the height of the barrier whose interaction with the barrier
can be examined independently due to above mentioned linearity
of the problem. In spite of the fact that the mean energy of the front $q$ is
larger than the height of the barrier, the plane waves in the above
superposition with energy less than the barrier height are completely reflected
and this reflection results in the above mentioned overshoot.
Next, we see the oscillating after-effect in the local density of the
reflected front (see the thick red curve in Fig.~\ref{fig3}(a)) that slowly
tends to the pole contribution with increasing of $\xi$ shown
by the thin black step-like curve. Comparing definitions \ref{dens}
and \ref{curr} and having in mind that both reflected wave components
differ only by sign ($v=-u$) we conclude that the same oscillations are present
in the current density as well. This after-effect is closely related to the
Zitterbewegung \cite{R24}. Usually this trembling motion of the electron is
explained by means of interference between positive- and negative-energy
relativistic wave function components. The interference itself is not
sufficient and some additional perturbation is necessary.
In our case this is the dispersive mass barrier that mixes the eigenfunctions of the
free electron motion. Compare with the reflection from the infinite barrier described
in Sec.~II where these components were not mixed.

\subsection{Below the barrier reflection ($q<1$)}

If the front energy is lower than the height of the barrier some
problem with the cut contribution appears because the pole is
located exactly in the interval $-1<x<1$ and interferes with the
integration contour $C_B$ that is shown in Fig.~\ref{fig2}(b).
Therefore, we changed the cut and transformed the
above integration contour into the $C_D$ as shown in
Fig.~\ref{fig2}(c). The pole contribution gives the same result as in
(\ref{rfpol}). Calculating the cut contribution as the
integral along the $C_D$ contour we changed the integration variable
$k=\exp[i(\pi+\ph)]$ in (\ref{reflpack}), presented this integral
as
\begin{equation}\label{rfpj2}
\begin{split}
  \Phi_{\mathrm{cut}}^{(\mathrm{rfl})}(\xi)
  = &-\frac{\sqrt{2}}{\pi}\int_0^{\pi/2} d\ph \sqrt{\cos\ph}\e^{-\xi\cos\ph}\\
  &\times\left(\frac{\e^{i(3\ph/2-\xi\sin\ph)}}{q + i\e^{i\ph}}
  +\frac{\e^{-i(3\ph/2-\xi\sin\ph)}}{q + i\e^{-i\ph}}\right)
  \end{split}
\end{equation}
and calculated this integral numerically. A typical result is shown in
Fig.~\ref{fig3}(b).

Comparing the curves in both Figs.~\ref{fig3} (a) and (b) we note some
differences. In the case of the below barrier reflection instead of an overshoot
in the red thick solid curve that corresponds to the leading
edge we see some diminishing of the electron density. This can be
explained by the same superposition of many Fourier harmonics as in the previous
above the barrier reflection case. Among them there are harmonics with energy larger
than the barrier height. They do not reflect but penetrate the barrier
what finally causes the above mentioned lack of density at the
reflected front.

\section{Penetration into the barrier}

The calculation of the front part that penetrates the barrier is
similar to the calculation of the reflected one presented in the
previous sections. The only difference is that the integral
(\ref{reflpack}) now is replaced by the integral (\ref{penpack}).
It leads to the different wave function components that depend on
both ($x$ and $t$) arguments.

Calculating the integral (\ref{penpack}) along the same contours
in the complex $k$-plane shown in Figs.~\ref{fig2}(a) and (b)
we divide this contribution into two parts. As in the case of reflection the pole
contribution is trivial:
\begin{equation}\label{trpole}
  \Psi_{\mathrm{pole}}^{\mathrm{(tr)}}(x,t) = \frac{\Theta(t-x)T(q)}{\sqrt{2q}}\left(
  \begin{array}{c}
  \sqrt{q+1} \\
  \sqrt{q-1} \\
  \end{array}
  \right)
  \e^{ix\sqrt{q^2-1}}\e^{-ikt}.
\end{equation}
It represents the result obtained by the scattering of plane
waves truncated at the point $x=t$ due to the finite velocity of the
front.

The contribution of the cut was calculated using the formulas
analogous to Eqs.~(\ref{rfpj1}) and (\ref{rfpj2}) depending on
whether the energy of the front is larger or smaller than the
height of the barrier.

\subsection{Above the barrier penetration ($q>1$)}

\begin{figure}[ht]
\begin{center}
\includegraphics[width=8cm]{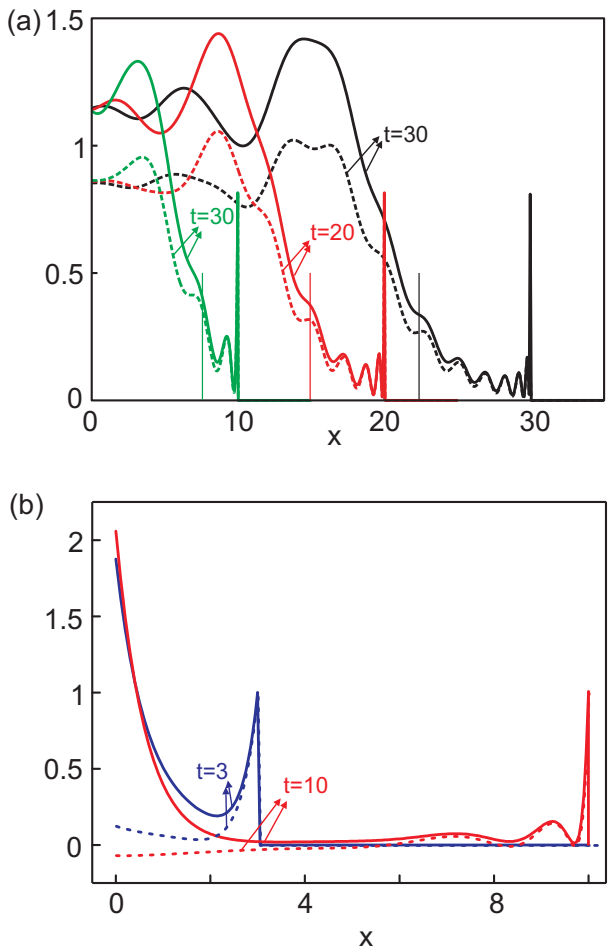}
\caption{(Color online) The local density (\ref{dens}) (solid curves)
  and current (\ref{curr}) (dotted curves) in the front that penetrates
  the barrier for different time $t$ values. The thin vertical lines
  indicate the motion with the group velocity $x=v_{\mathrm{gr}}t$:
  (a) -- above barrier reflection with $q=1.5$ and
  (b) -- below barrier reflection with $q=0.6$.} \label{fig4}
\end{center}
\end{figure}

A typical result for the penetration of the front into the barrier with
energy larger than the height of the barrier is shown in
Fig.~\ref{fig4}(a).
The local density and current is shown as a function of the
coordinate $x$ in the case of different time $t$ values.
As was already pointed out the barrier acts as a dispersive medium.
That is why the form of the front isn't conserved, and the local
density (actually the modulus of the wave function squared) and the
current demonstrates a rather sophisticated behavior. Three
important points are worth to be mentioned. First, the leading
edge of the front is transformed into a sharp peak that moves with
unit velocity inherent to the front in the dispersionless free
motion region ($x<0$). In the case of electromagnetic pulses it is known
as a precursor \cite{R14}. The physical
explanation of it's appearance is as follows. The theory of
dispersion in the classical \cite{jack62} and quantum \cite{zim64} version,
and the mechanical analogy of the Klein-Gordon equation \cite{ma99} as well, implies
that the medium in which the waves propagate has its own degrees of
freedom that actually causes the dispersion of waves in a
stationary regime. This stationary regime corresponds to the
excited state of the medium and, consequently, needs some time to be
established. That is why the first piece of the front (the
precursor itself) moves through the unexcited medium, and doesn't undergo
any dispersion. Speaking figuratively one can imagine that the precursor
prepares the media for the propagation of the main part of the pulse
or front. In an analogous way one may consider the mass barrier region
as a medium with inner degrees of freedom whose quantum analog is the
difference of the ground state energies of the two sublattices
when we consider the graphene in the tight binding approximation.
Consequently, the precursor has to be present in our Dirac electron case.

Second, as seen in Fig.~\ref{fig4}(a) the main part of the
front is retarded with respect to the above precursor. It caries
the energy of the front which is the reason why it moves with the group
velocity $v_{\mathrm{gr}}= d\eps/dq = \sqrt{q^2-1}/q$. In the case
of $q=1.5$ we have $v_{\mathrm{gr}}\approx 0.75$, which is
indicated by thin vertical lines in Fig.~\ref{fig4}(a).

And at last the third point is that in spite of the homogeneous
barrier region we see the waving behavior far from the leading
edge of the local density and current that are related to the
above mentioned zitterbewegung, and which is much more pronounced in
the penetrated front.

\subsection{Below the barrier penetration ($q<1$)}

A typical result is shown in Fig.~\ref{fig4}(b) for the case
of below barrier penetration when the front energy is lower than
the height of the barrier.

Here again we see the precursor moving with unit velocity
that makes the below the barrier reflection analogous to the above
the barrier one. But now instead of preparing the barrier for the
wave propagation this precursor constructs step by step the
evanescent wave that is inherent to the under the barrier
reflection of plane waves. This process needs some resources. That is
why the precursor loses its intensity (becomes narrower during its motion)
in contrary to the penetration with the energy larger than the height
of the barrier where such loses are not noticeable (compare the precursors
in Figs.~(a) and (b)). The oscillatory behavior of the penetrated fronts are also seen in the local density and current. This is in agreement  with the fact that the mass barrier is a dispersive medium.

\section{Penetration of the rectangular wave packet into the mass barrier}

\begin{figure}[ht]
\begin{center}
\includegraphics[width=8cm]{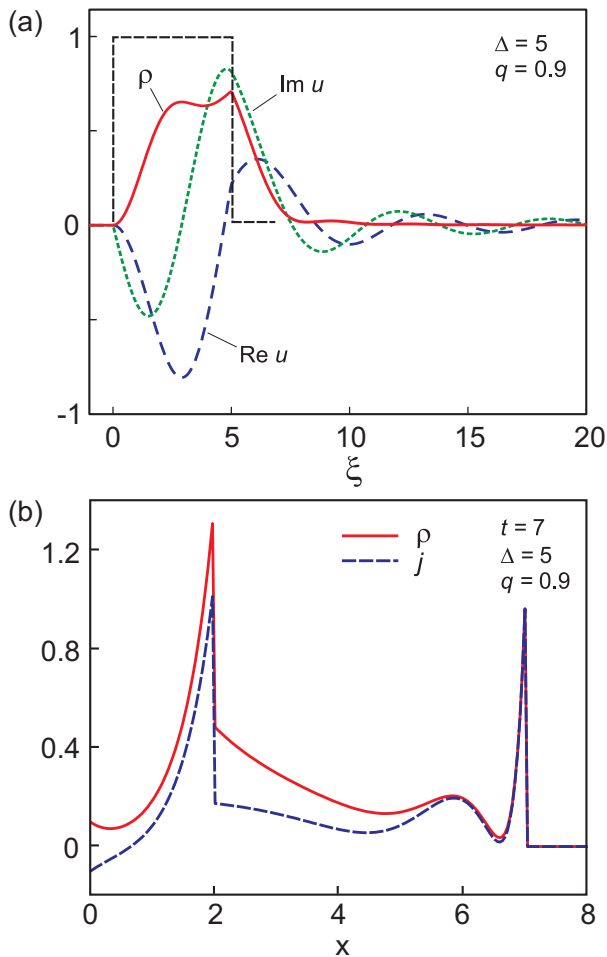}
\caption{(Color online) (a) -- the reflected rectangular wave packet
  with shift $\Delta=5$ between the opposite fronts
  as a function of $\xi=x+t$: the real (blue dashed curve),
  imaginary (green dotted curve) wave function components and the local
  density (red solid curve); black dashed
  rectangular shows the wave packet reflected by a hard wall.
  (b) -- the local density (solid curves) and current
  (dotted curves) for the penetrated part of the wave packet, at a given time.} \label{fig5}
\end{center}
\end{figure}

The reflection and penetration of the steep fronts considered in
the previous sections enables us to construct the result
for the rectangular wave packet reflection by the mass barrier. The
result is obtained as the superposition of two fronts with the proper
shift $\Delta$ and amplitudes chosen. The most interesting case is the below the
barrier reflection that is shown in Fig.~\ref{fig5} in the case of $q=0.9$.
The reflected wave packet is shown in part (a).
It depends on the argument $\xi=x+t$ as in the case of
the front considered in the previous sections.
The part of the wave packet that penetrates the barrier is a function of
both $x$ and $t$ arguments. The local density is shown as a function
of $x$ in Fig.~\ref{fig5}(b) for given time $t=7$ (red solid curve)
together with the current (blue dashed curve). Here we see two precursors.
One of them constructs the evanescent wave introducing some charge into
the barrier, and later the other one destroys it extracting that charge out
of the barrier region. The part of that evanescent wave is clearly seen between those precursors.
It is interesting to establish whether all charge is
extracted or some of it is left in the barrier and continues its motion
towards $+\infty$.

\begin{figure}[!ht]
  \begin{center}
  \includegraphics[width=8cm]{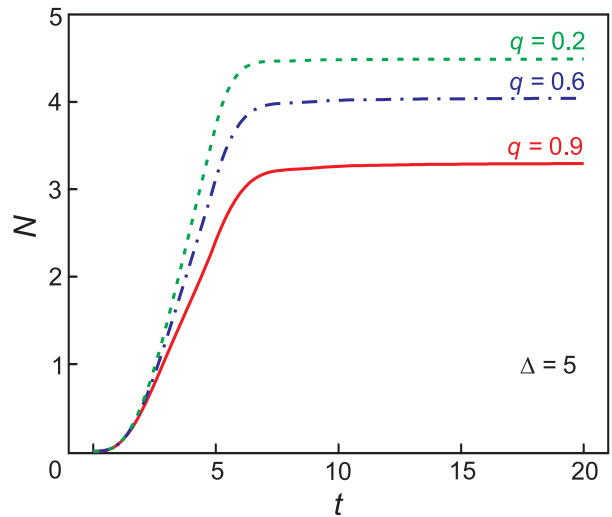}
  \end{center}
  \caption{(Color online) The number of particles in the reflected wave packet
  as a function of time for different $q$ values in the case when this number
  for incident wave packet is $N_0=5$.}
\label{fig6}
\end{figure}

The easiest way to clear this point up is to compare the density in the reflected wave packet
(the red solid curve in Fig.~\ref{fig5}(a)) with black dashed curve in
the same figure that represents the wave packet reflected by the hard wall.
It is evident that the area below the red solid curve is smaller than
the one below the black dashed curve. This means that the number of particles reflected by
the finite mass barrier is smaller than this amount in the incident wave packet, consequently,
some amount of the wave packet penetrates the barrier and moves there towards $+\infty$ in spite that
its mean energy is smaller than the height of the barrier ($q < 1$).
This fact is even better seen in Fig.~\ref{fig6} where the total number of particles
in the reflected wave packet is shown as a function of time.
Notice that $N$ is smaller than $N_0=5$ (the number of particles in the incident
wave packet) in asymptotic region (for large time $t$), indicating explicitly
that the closer $q$ is to unity the larger the amount of the wave packet
that penetrates the barrier.

In Fig.~\ref{fig5}(a) we see one more interesting feature. It is a long waving
tail that appears due to the above mentioned degrees of freedom of the barrier. They
are excited by the incident wave packet and then radiate the energy
even after the incident wave packet disappears. This is remarkable that mathematically this tail
follows from the contribution of the cut to the reflected wave function (\ref{rfpj2}).
In the case of large $\xi$ values (what corresponds to the large time $t$ asymptotic)
the contribution to the integral is given by the contour edges and leads to the standard
power type $1/\xi$ (non-exponential) behavior, that appears in many systems characterized
by a continuous spectrum, which for example were discovered in hydrodynamics \cite{ernst72}.

\section{Reflection and penetration of the Gaussian wave packet into the mass barrier}

Now let us consider the opposite case: the penetration into the
mass barrier of a wave packet with a rather soft envelope. We consider the
Gaussian wave packet replacing the initial condition (\ref{initinc}) by
the following one:
\begin{equation}\label{icgauss}
  \Psi^{(\mathrm{inc})}(x,0) = \frac{1}{\sqrt{2}}\left(
                                                   \begin{array}{c}
                                                     1 \\
                                                     1 \\
                                                   \end{array}
                                                 \right)
  \e^{iqx}\frac{\e^{-x^2/2\sigma}}{(\pi\sigma)^{1/4}}.
\end{equation}
Applying the same scheme as in section \ref{sec_rsf} we present
the above wave function as the Fourier integral (\ref{incfourier})
with the following Fourier transform:
\begin{equation}\label{ff}
  f(k) = (4\pi\sigma)^{1/4}\e^{-\sigma(q-k)^2/2}.
\end{equation}
Inserting this Fourier transform into Eq.~(\ref{incfourier})
and replacing the exponent $\exp(ikx)$ by the exponential solution of time
dependent Dirac equation (\ref{increfl},\ref{outgo}) as we did
before we obtain the following integral representations of the
incident Gaussian wave packet in the free motion region $-\infty < x < 0$
\begin{equation}\label{incpackg}
  \Psi^{(\mathrm{in})}(x,t) = \frac{\sigma^{1/4}}{2\pi^{3/4}}\left(
    \begin{array}{c}
      1 \\
      1 \\
    \end{array}\right)
  \int_C dk\e^{-\sigma(q-k)^2/2 + ik(x - t)},
\end{equation}
its reflected part in the same region
\begin{equation}\label{reflpackg}
\begin{split}
  \Psi^{(\mathrm{rfl})}(x,t) &= \frac{\sigma^{1/4}}{2\pi^{3/4}}\left(
    \begin{array}{c}
      1 \\
      -1 \\
    \end{array}\right) \\
  &\times\int_C dkR(k) \e^{-\sigma(q-k)^2/2 -ik(x + t)},
\end{split}
\end{equation}
and the one penetrated into the barrier ($0 < x < \infty$)
\begin{equation}\label{penpackg}
\begin{split}
  \Psi^{(\mathrm{tr})}(x,t) = &\frac{\sigma^{1/4}}{2\pi^{3/4}}
  \int_C dk \e^{-i R(k)x}\\
  &\times\left(
    \begin{array}{c}
      1 + R(k) \\
      1 - R(k) \\
    \end{array}\right)
  \e^{-\sigma(q-k)^2/2 + ik(x - t)}.
  \end{split}
\end{equation}

\begin{figure}[!ht]
  \begin{center}
  \includegraphics[width=6cm]{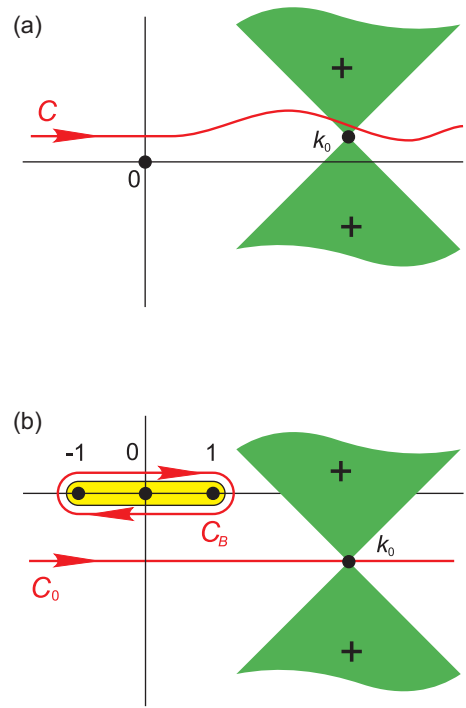}
  \end{center}
  \caption{(Color online) (a) -- Forbidden sectors in the complex
  $k$-plane and the integration contour for the calculation of the integrals
  (\ref{incpackg}-\ref{penpackg}). (b) -- Transformation of the integration
  contour in the case $x<t$.}
\label{fig7}
\end{figure}

The calculation of these three integrals differs, however, from the
calculation of the previous ones in the case of the steep front or
rectangular wave packet. The matter is that due to the exponent with
momentum $k$ squared the integrand has a more sophisticated singularity
at infinity. That is why the integration contour can not be
shifted to $\pm i\infty$, the method applied previously fails, and
we have to look for other possibilities to consider those
integrals. We demonstrate the method of calculation in the case of the most simple
integral (\ref{incpackg}) for the incoming wave packet. Analyzing the
argument of the single integrand exponent we see that the
complex $k$-plane can be divided into four sectors as shown
in Fig.~\ref{fig7}(a).
In two of them, shown by green shadowing, the exponent increase
for $|k|\to\infty$. Consequently, the integration contour $C$ has
to avoid these two sectors, and must be located completely in the
two other white sectors. It is evident that going from $-\infty$
to $\infty$ this integration contour has to cross the saddle point
$k_0$ that is obtained by equating the derivative of the argument
of the exponent to zero:
\begin{equation}\label{zeroder}
  \frac{d}{dk}\left[\sigma(k-q)^2/2 - ik(x - t)\right]\Big|_{k=k_0}
  = k_0 - q - i(x - t) = 0,
\end{equation}
and
\begin{equation}\label{saddle}
  k_0 = q + i(x - t).
\end{equation}
The value of the integral can be easily estimated by the saddle
point method. We used the numerical integration over the
contour $C_0$ shown in Fig.~\ref{fig7}(b) by the horizontal red solid
line. By the way the integral (\ref{incpackg}) can be calculated
analytically and it reads
\begin{equation}\label{incpacf}
  \Phi(x,t) = \frac{1}{\sqrt{2}}\left(
    \begin{array}{c}
      1 \\
      1 \\
    \end{array}\right)
  \e^{iq(x-t)}\frac{\e^{-(x-t)^2/2\sigma}}{(\pi\sigma)^{1/4}},
\end{equation}
what corresponds to the initial condition (\ref{icgauss}) that
moves with unit velocity to the right conserving its form, as it
should be in the dispersionless free motion region ($-\infty < x <
0$). Thus in the case of the Gaussian wave packet the saddle point
contribution to the integrals is the analog of the contribution of
the pole in the case of a rectangular wave packet (or in the pure
exponent case).

Integrals (\ref{reflpackg}) and (\ref{penpackg}) are more
complicated due to the radical in the function $R(k)$ in the
integrands. Because of these radicals there are branching points,
and we have to make the cut in the complex $k$-plane connecting
them as it is shown in Fig.~\ref{fig7}(b).
Calculating these integrals it is important to check
the relative position of the cut and the saddle point that
according to (\ref{saddle}) depends on the sign of
$(x-t)$. So, if $x>t$ the horizontal contour $C_0$ crossing the saddle
point lays above the cut, and consequently, only that contribution
has to be taken into account. In the case of $x<t$ as we see in
Fig.~\ref{fig7}(b) the contour $C_{0}$ is located below the cut.
In this case the contribution of the cut
(namely, the integration over the contour $C_B$) has to be added.
In the case of integrals (\ref{reflpackg}) and (\ref{penpackg}) we
performed the integration numerically taking into account the
above mentioned features of the contours.

\begin{figure}[!ht]
  \begin{center}
  \includegraphics[width=8cm]{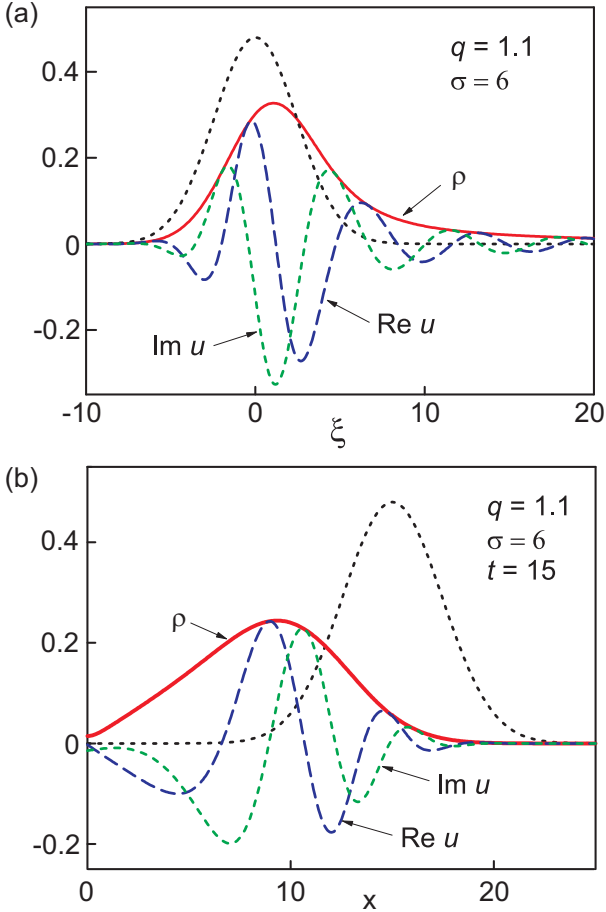}
  \end{center}
  \caption{(Color online) Above the barrier reflection of the Gaussian wave packet:
  the blue dashed and green dotted curves -- the real and imaginary part of wave function
  component $u$ (for reflected wave packet it coincides with the component $-v$),
  and the red solid curve -- the density.
  (a) -- reflected wave packet as a function of $\xi=x+t$,
  the thin dashed black curve -- the density of the incident wave packet
  when it is reflected from the hard wall as in Sec.~\ref{sec_hw}.
  (b) -- The wave packet that penetrates the barrier,
  the thin dashed black curve -- the density of the incident wave packet
  propagating in the absence of the barrier.}
  \label{fig8}
\end{figure}

\begin{figure}[!ht]
  \begin{center}
  \includegraphics[width=8cm]{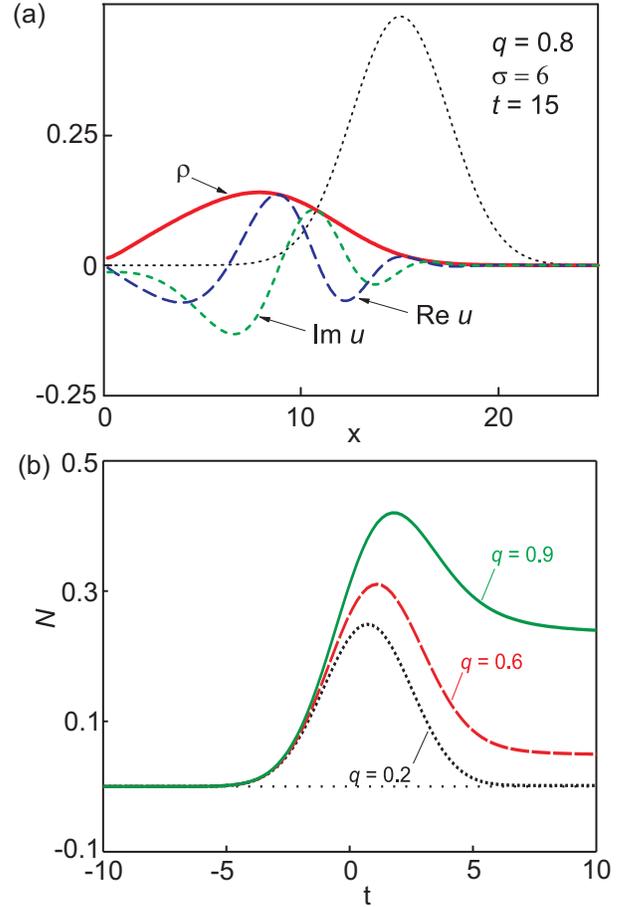}
  \end{center}
  \caption{(Color online) ({\it a}) -- the same as in \ref{fig8}({\it b})
  but for the below the barrier reflection case;
  ({\it b}) -- the total number of particles in the barrier (\ref{number}).}
\label{fig9}
\end{figure}

A typical result for the above barrier reflection of the Gaussian wave
packet is shown in Fig.~\ref{fig8}.
In part (a) the reflected wave packet is shown. We see that it
differs form the one of the rectangular wave packet or front (see
for comparison Figs.~\ref{fig3} or \ref{fig5}(a)). There is no
overshoot neither a lack of intensity at the edges, and the
reflected wave packet conserves more or less its Gaussian form.
The matter is that the Gaussian wave packet actually has no
leading and trailing edges. The increase (or decrease) of
intensity in the Gaussian wave packet is slow, the barrier manages
to adjust itself for reflection, and the sophisticated features
that we met in the case of rectangular wave packet reflection are
not present. The single reminder of the previously considered
reflection of the rectangular wave packet is the small asymmetry
of the reflected Gaussian wave packet, some retardation of it as
compared with the reflection of the Gaussian wave packet by the
hard wall (shown by the thin dashed black curve), and the long
waving tail after the reflected wave packet. The wave packet that
penetrates the barrier is shown in Fig.~\ref{fig8}(b). This
differs qualitatively from the rectangular wave packet case. There
is no precursors, and the form of the penetrated wave packet is
rather close to a Gaussian. It is remarkable that this form is
more or less the same even in the case of below the barrier
reflection ($q<1$) as seen in Fig.~\ref{fig9}(a), only the
amplitude being smaller and the width larger what is caused by the
dispersion of the barrier medium.
We see no precursors and no formation of an evanescent wave. So,
in the case of the below the barrier reflection the penetrated
wave packet exhibits qualitatively the same Gaussian form with
small asymmetry and this packet is essentially retarded as
compared with the motion of the wave packet in the absence of the
barrier (shown by the thin dashed black curve), and a rather large
spreading that, as it was already mentioned, appears due to the
wave dispersion in the barrier. To our mind the retardation of the
penetrated wave packet apparently illustrates the problem of the
time interval that the wave packet spends in the barrier that was
intensively discussed 15 -- 20 years ago (see the discussion in
Ref.~\onlinecite{hauge89}).

In the case of the below the barrier penetration the part of the Gaussian
wave packet continues its motion in the barrier towards $+\infty$ as it
was in the rectangular wave packet case, what is evidently seen in
Fig.~\ref{fig9}(b) where the total number of particles
in the barrier is shown as a function of time $t$. This is clearly demonstrated by the non-zero asymptotic
of these values at large $t$ explicitly indicate that. These values
increase when the energy $q$ approaches the top of the barrier being
essentially larger for short wave packets.

\section{Conclusions}

We considered a simple 1D model of wave packet reflection and
penetration into a mass barrier. Interesting behavior was already
found when considering reflection off a hard wall which is due
to the fact that the wave packet isn't a point particle.
For example we found already reflection before the wave packet reached the
wall, there was a compression of the wave packet during reflection and the
width of the wave packet was restored after reflection.

The dependence on the wave packet shape was discussed
by considering two limiting shapes, namely, the rectangular wave packet
with sharp leading and trailing edges, and the Gaussian wave packet
which has a rather soft form.

The most crucial differences of the wave packet reflection and
penetration into the mass barrier with those calculated for
standard scattering of exponents were revealed in the case of
the rectangular wave packet. Here the dispersion of the electron wave
in the barrier showed itself to the full extent. We found
different structures moving with different velocities. The leading
edge of the front moves with velocity equal to unity because
it moves through the unprepared dispersive medium of the barrier.
Meanwhile the main piece of the wave packet (or front) moves, however,
with the group velocity which is the reason why it is retarded.

At the leading edge of the reflected front we see some overshoot
or failure of some intensity due to the fact that the front itself
is a superposition of many exponents that are reflected with
different probabilities. In the case of a soft Gaussian wave packet the
features of reflection and penetration are quite different. The
main difference is the absence of the precursors. The matter is
that the soft leading edge of the wave packet prepares the medium of the
dispersive barrier gradually, and there is no possibility for
the precursor (moving in the unprepared media) to appear.
Nevertheless some rudiments of the above features can still be
revealed. That is the distortion of the wave packet form, making its
leading edge sharper and the trailing edge more prolonged. Next,
we see some retardation of the reflected and penetrated wave packet as
compared with the motion in the absence of the barrier.

We also demonstrated that for the Gaussian wave packet in the case when the mean wave packet energy is lower than the height of the barrier part of the wave packet penetrates into the barrier and moves along it, more or
less conserving its shape.

\section{Acknowledgment}

This research was supported by the Flemish Science Foundation
(FWO-Vl) and (in part) by the Lithuanian Science Council under project
No.~MIP-79/2010.

\end{document}